\documentstyle[aps,epsfig,multicol]{revtex}

\renewcommand{\vec}{\bf} % USE BOLDFACE FOR 3-VECTOR, DO NOT DELETE OR CHANGE THIS LINE %

\begin{document}
\draft
\title{Enhanced photon production from quark-gluon plasma: Finite-lifetime effect}
\author{Shang-Yung Wang\thanks{E-mail: sywang@phyast.pitt.edu}
and Daniel Boyanovsky\thanks{E-mail: boyan@pitt.edu}}
\address
{Department of Physics and Astronomy, University of Pittsburgh, Pittsburgh,
Pennsylvania 15260}
\date{December 18, 2000}
\maketitle

\begin{abstract}
Photon production from a thermalized quark-gluon plasma of finite lifetime is
studied directly in real time with a nonequilibrium formulation that includes
off-shell (energy nonconserving) effects. To lowest order we find that
production of direct photons form a quark-gluon plasma of temperature $T\sim
200\;{\rm MeV}$ and lifetime $t\sim 10-20\;{\rm fm}/c$ is strongly enhanced by
off-shell (anti)quark bremsstrahlung $q(\bar{q})\rightarrow q(\bar{q})\gamma$.
The yield from this nonequilibrium finite-lifetime effect dominates over those
obtained from higher order equilibrium rate calculations in the range of
energy $E>2\;{\rm GeV}$ and falls off with a power law for $E \gg T$.
\end{abstract}


\begin{multicols}{2}

The observation of a novel phase of matter, the quark-gluon plasma (QGP), is
one of the most important goals of ultrarelativistic heavy ion experiments
currently undertaken at CERN SPS and BNL RHIC~\cite{cern}. The quark-gluon
plasma formed in the early stage of the collision expands and cools rapidly to
a mixed phase of quarks, gluons, and hadrons, and ultimately undergoes a
freeze-out from a state of hadronic gas. Estimates based on energy deposited
in the central collision region at RHIC energies $\sqrt{s}\sim 200A\;{\rm
GeV}$ suggest that the lifetime of a deconfined phase of quark-gluon plasma is
of order $10-20\;{\rm fm}/c$ with an overall freeze-out time of order
$100\;{\rm fm}/c$~\cite{qgp}. An important aspect is an assessment of
nonequilibrium effects associated with the rapid expansion and finite lifetime
of the plasma and their impact on experimental observables.

Amongst various experimental signatures proposed to detect the
quark-gluon plasma phase,
photons (both real and virtual) have long been considered as the most
promising direct signals~\cite{Shu78}. This is because,
unlike strongly interacting hadrons, photons have a mean free
path much larger than the typical size
of the plasma formed in ultrarelativistic heavy ion collisions.
Once produced they  escape  from the system
without further interaction, thus carrying clean
information from the early hot quark-gluon plasma phase.

The first observation of direct photon production in ultrarelativistic heavy
ion collisions has been recently reported by the CERN WA98 Collaboration in
Pb+Pb collisions at $\sqrt{s}= 158A\;{\rm GeV}$~\cite{WA98}. The transverse
momentum distribution of direct photons is determined on a statistical basis
and compared to the background photon yield predicted from a calculation of the
radiative decays of hadrons. The most interesting result is that a significant
excess of direct photons beyond that expected from proton-induced reaction at
the same $\sqrt{s}\,$ is observed in the range of transverse momentum greater
than about $1.5\;{\rm GeV}/c$ in central collisions. While it is not yet clear
whether a QGP was formed in the central collision region at SPS energies, this
result does suggest the experimental feasibility of direct photon production as
a signal of the QGP phase, expected to be formed at RHIC energies.

One goal of this article is to study {\em directly in real time} the effect of
the finite QGP lifetime on direct photon production. We focus on the direct
photon yield from a thermalized quark-gluon plasma of temperature $T\sim
200\;{\rm MeV}$ and lifetime $t\sim 10-20\;{\rm fm}/c$ in accordance with
estimates based on collision energies reached at RHIC. Another goal is to
compare this nonequilibrium photon yield to those of previous
investigations~\cite{kapusta,aurenche} that obtain an equilibrium production
rate by taking the thermal average of transition amplitudes, which in effect
is tantamount to assuming an infinite lifetime for the QGP. It is {\em not}
the purpose of this article to compare direct photon production from the QGP
with that from the hadronic gas, as this has been studied in detail in
Refs.~\cite{kapusta,aurenche,hatsuda}.

Many investigations in the literature have been devoted to hard real photon
production from the quark-gluon plasma~\cite{kapusta,aurenche,hatsuda,baier1}.
Assuming a QGP in thermal equilibrium with an {\em infinite} lifetime, these
authors focused on the equilibrium photon production rate, which for photons
of momentum ${\vec p}$ is given by~\cite{kapusta,aurenche,baier1,lebellac}
\begin{equation}
E\frac{dN}{d^3p\,d^4x}= -\frac{2}{(2\pi)^3}\,n_B(E)\,{\rm Im}
\Pi^{R}(E).\label{rate}
\end{equation}
Here $E=|{\vec p}|$, $\Pi^{R}(E)$ is the retarded transverse photon self-energy
at finite temperature $T$ evaluated on shell, and $n_B(E)=1/(e^{E/T}-1)$ is the
Bose-Einstein distribution. Using the hard thermal loop (HTL) resummed
effective perturbation theory developed by Braaten and Pisarski~\cite{htl},
Kapusta {\it et al.}~\cite{kapusta} and Baier {\it et al.}~\cite{baier1} showed
that at one-loop order (in effective perturbation theory) the processes that
contribute to photon production are the gluon-to-photon Compton scattering off
(anti)quark $q(\bar{q})g\rightarrow q(\bar{q})\gamma$ and quark-antiquark
annihilation to photon and gluon $q\bar{q}\rightarrow g\gamma $. The
corresponding rate of energetic ($E\gg T$) photon emission for two light quark
flavors ($u$ and $d$ quarks) is given by~\cite{kapusta}
\begin{equation}
E\frac{dN}{d^3p\,d^4x}\bigg|_{\text{one-loop}}
=\frac{5}{9}\frac{\alpha\alpha_s}{2\pi^2}T^2 e^{-E/T}
\ln\left(\frac{0.23 E}{\alpha_s T}\right),\label{rate:kapusta}
\end{equation}
where $\alpha$ is the fine-structure constant and $\alpha_s=g_s^2/4\pi$ with
$g_s$ being the strong coupling constant. In a recent development Aurenche {\it
et al.}~\cite{aurenche} have found that the two-loop contributions to the
photon production rate arising from (anti)quark bremsstrahlung
$qq(g)\rightarrow qq(g)\gamma$ and quark-antiquark annihilation with scattering
$q\bar{q}q(g)\rightarrow q(g)\gamma$ are of the same order as those evaluated
at one loop. The two-loop contributions to the photon production rate
read~\cite{aurenche}
\begin{eqnarray}
E\frac{dN}{d^3p\,d^4x}\bigg|_{\text{two-loop}}
&=&\frac{40}{9}\frac{\alpha\alpha_s}{\pi^5}T^2 e^{-E/T}
(J_T-J_L)\nonumber\\
&&\times\left[\ln 2+\frac{E}{3T}\right],\label{rate:aurenche}
\end{eqnarray}
where $J_T\approx 4.45$ and $J_L\approx -4.26$ for two light quark flavors.
Most importantly, they showed that the two-loop contributions completely
dominate the photon emission rate at high photon energies~\cite{aurenche}.

To study photon production from a QGP of {\em finite lifetime}, we use a
real-time kinetic approach~\cite{boyankin} based on nonequilibrium quantum
field theory~\cite{CTP,boyanrgk,boyanQED}, which when improved by a resummation
via a dynamical renormalization group provides a consistent microscopic
derivation of quantum kinetics from the underlying
theories~\cite{boyanrgk,boyanQED}. One of the advantages of this real-time
kinetic approach is that it is capable of capturing off-shell (energy
nonconserving) effects originating in the finite system lifetime, as {\em
completed collisions} are not assumed {\it a priori}~\cite{boyanQED}.

Because of the abelian nature of the electromagnetic interaction, we will work
in a gauge invariant formulation in which physical observables (in the
electromagnetic sector) are manifestly gauge
invariant~\cite{boyanrgk,boyanQED}. Since photons escape directly from the
quark-gluon plasma without further interaction, it is adequate to treat them
as asymptotic particles. In the Heisenberg picture the number operator ${\rm
N}({\vec p},t)$ that counts the total number of photons of momentum ${\vec p}$
(per phase space volume) at time $t$ is defined by
\begin{eqnarray}
{\rm N}({\vec p},t)&=&\sum^2_{\lambda=1}a^\dagger_\lambda({\vec p},t)
a_\lambda({\vec p},t),
\end{eqnarray}
where $a_\lambda({\vec p},t)$ [$a^\dagger_\lambda({\vec p},t)$] is the
annihilation (creation) operator that destroys (creates)  a photon of momentum
${\vec p}$ and polarization $\lambda$ at time $t$. The {\em time-dependent}
photon production rate, which up to a trivial factor $E/(2\pi)^3$ is related to
the expectation value of the time derivative of ${\rm N}({\vec p},t)$, can be
obtained by using the Heisenberg equations of motion (for details, see
Ref.~\cite{boyanQED}). In the framework of nonequilibrium quantum field theory,
the time-dependent photon production rate is given by~\cite{boyanQED}
\begin{eqnarray}
E\frac{dN(t)}{d^3p\,d^4x} &=&\lim_{t'\rightarrow t}\sum_{\text{\scriptsize
flavors}} \frac{3\,e_q}{2(2\pi)^3} \left(\frac{\partial}{\partial
t'}-iE\right) \int\frac{d^3q}{(2\pi)^3}\nonumber\\ &&\times\; \big\langle
\bar{\psi}^-(-{\vec k},t){\bbox\gamma} \cdot{\vec A}^+_T({\vec
p},t')\psi^-({\vec q},t) \big\rangle+ {\rm c.c.},\nonumber\\\label{rate1}
\end{eqnarray}
where ${\vec k}={\vec p}+{\vec q}$. Here $e_q$ is the electromagnetic coupling
constants of the quarks, and $\psi$ denotes the (gauge invariant) quark field.
The ``$+$'' (``$-$'') superscripts for the fields refer to fields defined in
the forward (backward) time branch~\cite{CTP}.

As usual we assume that there are no photons initially and those once produced
will escape from the plasma without building up their population. Therefore the
QGP in effect is treated as the vacuum for the photons. Consequently, the
nonequilibrium expectation values on the right-hand side of Eq.~(\ref{rate1})
is computed perturbatively to order $\alpha$ and in principle to all orders in
$\alpha_s$ by using real-time Feynman rules and propagators. Furthermore, the
photon propagators are the same as those in the vacuum.

We shall further assume the weak coupling limit $\alpha \ll \alpha_s \ll 1$.
Whereas the first limit is justified and essential to the interpretation of
electromagnetic signatures as clean probes of the QGP, the second limit can
only be justified for very high temperatures, and its validity in the regime of
interest can only be assumed so as to lead to a controlled perturbative
expansion.

It is not hard to see that the lowest order contribution to Eq.~(\ref{rate1})
is of one quark loop and of order $\alpha$. In using bare quark propagators we
consider the quark momentum in the loop to be hard, i.e., $q\gtrsim T$. Soft
quark lines require HTL resummed effective quark propagator leading to higher
order corrections. Indeed, the one-loop diagram with soft quark loop momentum
is part of the higher order contribution of order $\alpha\alpha_s$ that has
been calculated in Refs.~\cite{kapusta,baier1}.

In this article we focus on the {\em lowest} order contribution which, as will
be understood below, contributes to direct photon production solely as a
consequence of the finite QGP lifetime, and has therefore been missed by all
previous investigations which assumed an infinite QGP lifetime.

For two light flavors ($u$ and $d$ quarks), the lowest order time-dependent
photon production rate is found to be given by~\cite{boyanQED}
\begin{equation}
E\frac{dN(t)}{d^3p\,d^4x}=\frac{2}{(2\pi)^3}
\int_{-\infty}^{+\infty}d\omega\,{\cal R}(\omega)\,
\frac{\sin[(\omega-E)(t-t_0)]}{\pi(\omega-E)}, \label{rate2}
\end{equation}
with
\begin{eqnarray}
{\cal R}(\omega)&=&\frac{20\,\pi^2\alpha}{3} \int\frac{d^3q}{(2\pi)^{3}}
\Big\{2\big[1-({\vec\hat{p}}\cdot{\vec\hat{k}})({\vec\hat{p}}
\cdot{\vec\hat{q}})\big]\nonumber\\
&&\times\,n_F(k)[1-n_F(q)]\delta(\omega-k+q)\nonumber\\ &&
+\,\big[1+({\vec\hat{p}}\cdot{\vec\hat{k}})({\vec\hat{p}}
\cdot{\vec\hat{q}})\big]\big\{n_F(k)n_F(q)\nonumber\\
&&\times\,\delta(\omega-k-q)+[1-n_F(k)]\nonumber\\
&&\times\,[1-n_F(q)]\delta(\omega+k+q)\big\}\Big\}, \label{R1}
\end{eqnarray}
where $t_0$ is the initial time at which the QGP is formed, $q=|{\vec q}|$, and
$n_F(q)=1/(e^{q/T}+1)$ is the Fermi-Dirac distribution.

A detailed analysis shows that the first delta function $\delta(\omega-k+q)$
with support below the light cone ($\omega^2<E^2$) corresponds to the Landau
damping cut, and the last two delta functions $\delta(\omega\mp k\mp q)$ with
supports above the light cone correspond to the usual two-particle cut.
Furthermore, one recognizes that ${\cal R}(\omega)$ has a physical
interpretation in terms of the following {\em off-shell} (energy
nonconserving) photon production processes: the first term describes
(anti)quark bremsstrahlung $q(\bar{q})\rightarrow q(\bar{q})\gamma$; the
second term describes quark-antiquark annihilation to photon
$q\bar{q}\rightarrow\gamma$; and the third term describes creation of a photon
and a quark-antiquark pair out of the vacuum $0\rightarrow q\bar{q}\gamma$. We
would like to stress that the photon and (anti)quark in these processes are
all on-shell particles, and off-shellness refers to nonconservation of energy
as will be explained below. It is also noted that the first two processes are
inherently distinct from the third one. Whereas the first two processes arise
solely due to the presence of the QGP, the third one in effect is a ``vacuum''
process that is Pauli suppressed during the QGP lifetime.

It can be shown that ${\cal R}(\omega)$ is related to the imaginary part of the
one-loop retarded transverse photon self-energy with hard loop
momentum~\cite{boyanQED}
\begin{equation}
{\cal R}(\omega)=-n_B(\omega)\,{\rm Im}\Pi^{R}(\omega)|_{\text
{one-loop}}.\label{R2}
\end{equation}
If, as is usually done, the QGP is assumed to be of infinite lifetime, then we
can take the infinite time limit $t_0\rightarrow -\infty$ in the argument of
the sine function in Eq.~(\ref{rate2}) and use the approximation
\begin{equation}
\frac{\sin[(\omega-E)t]}{\pi(\omega-E)}\buildrel{t\rightarrow
\infty}\over{\approx}\delta(\omega-E).
\end{equation}
This is the assumption of completed collisions that is invoked in
time-dependent perturbation theory leading to Fermi's golden rule and energy
conservation. Under this assumption  Eq.~(\ref{rate}) is recovered to lowest
order and one finds a {\em time-independent} photon production rate
proportional to ${\cal R}(E)$, provided that the latter is finite.

In the present situation, however, the three delta functions in ${\cal
R}(\omega)$ cannot be satisfied on the photon mass shell. Therefore under the
assumption of completed collisions the off-shell contribution to the photon
production rate simply vanishes due to kinematics. Physically this reflects
the energy nonconserving nature of the corresponding photon production
processes. The ``vacuum'' process is independent of the presence of the QGP or
its lifetime and persists for a infinitely long time. Therefore for the {\em
third} term of ${\cal R}(\omega)$ we have to take the infinite time limit,
which leads to the vanishing of the off-shell ``vacuum'' process $0\rightarrow
q\bar{q}\gamma$ by energy conservation. Only the ``medium'' processes depend
on the presence and finite lifetime of the QGP, hence we only need to consider
off-shell (anti)quark bremsstrahlung and quark-antiquark annihilation in the
rest of our discussion.

For any finite QGP lifetime the time-dependent rate given in Eq.~(\ref{rate2})
is finite and nonvanishing, thus leading to a nontrivial contribution to direct
photon production. In this article we focus on the photon yield instead of the
photon production rate, since the former is the quantity of phenomenological
interest that can be measured in experiments. The photon yield is obtained by
integrating the rate over the lifetime of the QGP. Using Eq.~(\ref{rate2}), one
obtains
\begin{equation}
E\frac{dN(t)}{d^3p\,d^3x}=\frac{2}{(2\pi)^3}
\int_{-\infty}^{+\infty}d\omega\,{\cal R}(\omega)\,
\frac{1-\cos[(\omega-E)t]}{\pi(\omega-E)^2},\label{yield}
\end{equation}
where, here and henceforth, we have set $t_0=0$.

Before proceeding to a numerical study, we give an analytic estimate of the
behavior of the photon yield in the HTL approximation. In this approximation
the leading contribution to ${\rm Im}\Pi^{R}(\omega)|_{\text{one-loop}}$ is
dominated by the Landau damping cut~\cite{lebellac,boyanQED}, which
corresponds to off-shell (anti)quark bremsstrahlung. Thus, from
Eq.~(\ref{R2}), we find~\cite{boyanQED}
\begin{equation}
{\cal R}_{\rm HTL}(\omega)=\frac{20}{3}\frac{\pi^2 \alpha T^2}{12}
\frac{\omega}{E}\left(1-\frac{\omega^2}{E^2}\right)n_B(\omega)\,
\theta(E^2-\omega^2).\label{RHTL}
\end{equation}
For $E\ll T$, ${\cal R}_{\rm HTL}(\omega)$ can be further simplified as
\begin{equation}
{\cal R}_{\rm HTL}(\omega)\buildrel{E\ll T}\over\simeq
\frac{20}{3}\frac{\pi^2 \alpha T^3}{12E}
\left(1-\frac{\omega^2}{E^2}\right)
\theta(E^2-\omega^2).\label{RHTL2}
\end{equation}
The dominant contribution of the $\omega$-integral in Eq.~(\ref{yield}) for
$E\ll T$ arises from the region where the resonant denominator vanishes, i.e.,
$\omega\approx E$~\cite{boyanrgk,boyanQED}. Using Eq.~(\ref{RHTL2}), we
obtain~\cite{boyanQED}
\begin{equation}
E\frac{dN(t)}{d^3p\,d^3x}\buildrel{E\ll T}\over=
\frac{5}{9}\frac{\alpha}{2\pi^2}\frac{T^3}{E^2}
\left[\ln 2Et+\gamma-1\right]+{\cal O}\left(\frac{1}{t}\right),
\label{yieldhtl}
\end{equation}
where $\gamma=0.577\ldots$ is the Euler-Mascheroni constant. It has been shown
in Ref.~\cite{boyanQED} that for photons of energy $E\lesssim T$ the
finite-lifetime contribution to the photon yield, Eq.~(\ref{yieldhtl}), is
{\em comparable at early times} to that of order $\alpha\alpha_s$ obtained
from the equilibrium rate given in Eq.~(\ref{rate:kapusta}).

\begin{center}
\begin{figure}
\epsfig{file=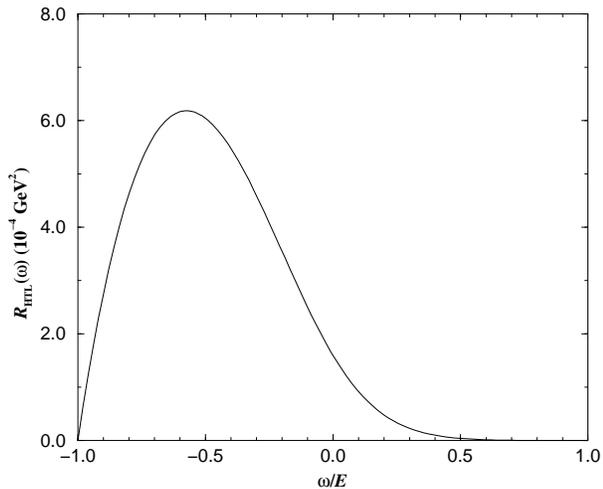,width=3.1in,height=2.6in} \vspace{.2in} \caption{The
function ${\cal R}_{\rm HTL}(\omega)$ is plotted in the limit $E\gg T$. Here
we take $E=2\;{\rm GeV}$ and $T=200\;{\rm MeV}$.} \label{fig:RHTL}
\end{figure}
\end{center}

Fig.~\ref{fig:RHTL} shows that for $E\gg T$, ${\cal R}_{\rm HTL}(\omega)$ is
exponentially suppressed in the region $T<\omega<E$. From this observation we
emphasize that (i) because for $E\gg T$ the threshold contribution near the
photon mass shell $\omega\approx E$ is exponentially suppressed, the
$\omega$-integral in Eq.~(\ref{yield}) is now dominated by the interval
$-E<\omega<T$, which corresponds to highly off-shell (anti)quark
bremsstrahlung; (ii) as the integrand in Eq.~(\ref{yield}) is
positive-definite and the function $1-\cos[(\omega-E)t]$ averages to 1 for
large $t$ in the region $-E<\omega<T$, for {\em fixed} $E\gg T$ the yield
approaches a constant at large times; and (iii) in contrast to those obtained
from equilibrium rates, the yield for $E\gg T$ is {\em not} suppressed by the
Boltzmann factor $e^{-E/T}$. These important nonequilibrium aspects have
noteworthy phenomenological implications studied in detail below.

We now perform a numerical analysis of the nonequilibrium finite-lifetime
contribution to photon production directly in terms of the yield given by
Eqs.~(\ref{yield}) and (\ref{R1}) [but, as explained above, without including
the third term of ${\cal R}(\omega)$] and compare the results to those
obtained from the equilibrium rates given in Eqs.~(\ref{rate:kapusta}) and
(\ref{rate:aurenche}). For this we consider a QGP of temperature $T=200\;{\rm
MeV}$ and lifetime $t=10-20\;{\rm fm}/c$. As remarked above, this is
approximately the scales of the QGP temperature and lifetime expected at RHIC
energies. For the value of $\alpha_s$ at this temperature, we use
$\alpha_s(T)=6\pi/(33-2N_f)\ln(8T/T_c)$~\cite{karsch}, where $N_f=2$ is the
number of quark flavors and $T_c$ is the quark-hadron transition temperature.
Taking $T_c=160\;{\rm MeV}$ from lattice QCD calculation for two quark
flavors, we find $\alpha_s\approx 0.3$ at $T=200\;{\rm MeV}$.

\begin{center}
\begin{figure}
\epsfig{file=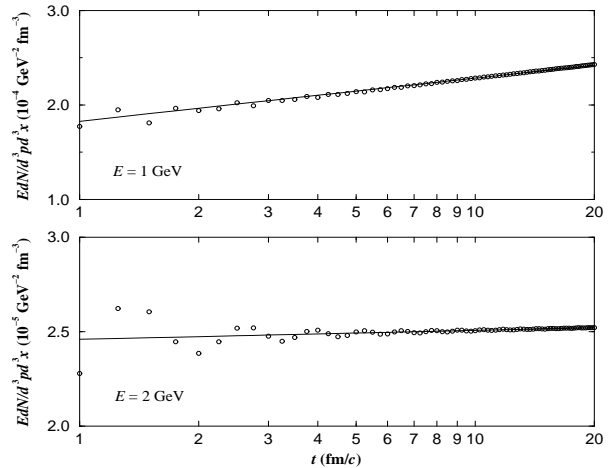,width=3.1in,height=2.6in} \vspace{.2in} \caption{The
nonequilibrium yield from a QGP of temperature $T=200\;{\rm MeV}$ is plotted
as a function of time for $E=1$ (upper) and $2\;{\rm GeV}$ (lower). The
circles denote the numerical result and the solid line is a logarithmic
fit.}\label{fig:yield}
\end{figure}
\end{center}

\begin{center}
\begin{figure}
\epsfig{file=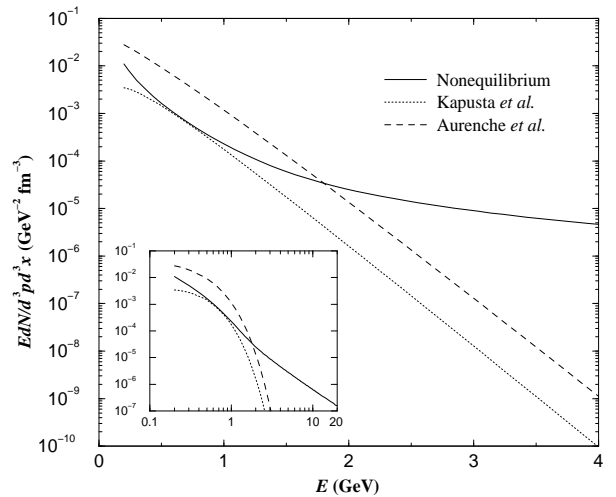,width=3.1in,height=2.6in} \vspace{.2in}
\caption{Comparison of various contributions to the direct photon yield from a
quark-gluon plasma of temperature $T=200\;{\rm MeV}$ and lifetime $t=10\;{\rm
fm/}c$. The inset shows the figure on a log-log plot.} \label{fig:spectrum}
\end{figure}
\end{center}

The results of the numerical study of the nonequilibrium photon yield are
displayed in Figs.~\ref{fig:yield} and \ref{fig:spectrum}. Fig.~\ref{fig:yield}
depicts the time evolution of the nonequilibrium yield up to $20\;{\rm fm}/c$
for photons of energies $E=1$ and $2\;{\rm GeV}$. We observe that the yield
exhibits clearly a logarithmic time-dependence at late times during the QGP
lifetime. Numerical evidence shows that the nonequilibrium yield $E
dN(t)/d^3p\,d^3x$ falls off with a power law $E^{-\nu}$ with $\nu\approx 2.14$
for $E\gg T$. In addition, the numerical result also reveals that the dominant
photon production process is (anti)quark bremsstrahlung $q(\bar{q})\rightarrow
q(\bar{q})\gamma$.

Figure~\ref{fig:spectrum} shows a comparison of the nonequilibrium and
equilibrium (one-loop and two-loop) contributions to the direct photon yield
in the range of energy $T < E < 4\;{\rm GeV}$. Whereas the two-loop
contribution dominates the direct photon yield for smaller values of $E$, a
significant enhancement of the direct photon yield due to the nonequilibrium
contribution is seen at $E>2\;{\rm GeV}$ as a consequence of its power law
falloff for $E\gg T$. In particular, the nonequilibrium contribution is larger
than the equilibrium contributions by several orders of magnitude for
$E>3\;{\rm GeV}$. As the nonequilibrium contribution for fixed $E\gg T$
approaches a constant at large times, the equilibrium contributions, which
grow linearly in time, will eventually dominate the yield if the QGP has a
very long lifetime. However, the linear growth in time of the equilibrium
contributions has to compensate the Boltzmann suppression for $E\gg T$.
Therefore we emphasize that for $E\gg T$ the equilibrium contributions {\em
could} dominate the yield {\em if} the QGP lifetime is of order
$10^2-10^3\;{\rm fm}/c$ or larger, which nevertheless is very unrealistic at
RHIC energies.

Our analysis can be extended to the case of direct photon production from a
QGP away from chemical equilibrium by replacing the equilibrium (anti)quark
distributions with the undersaturated ones~\cite{baier2}. We expect the main
features to remain in the more general situation.

In this article we have studied directly in real time the production of direct
photons from a thermalized QGP of temperature $T\sim 200\;{\rm MeV}$ and
lifetime $t\sim 10-20\;{\rm fm}/c$. To lowest order in perturbation theory we
find that direct photon production features a power law spectrum for $E\gg T$
due to a significant enhancement by off-shell (anti)quark bremsstrahlung.

Our main conclusion is that the novel nonequilibrium effect originating in the
finite lifetime of a thermalized QGP leads to an enhancement in the direct
photon yield for $E>2\;{\rm GeV}$, which dominates over the yields obtained
from equilibrium rate calculations for the same QGP lifetime. To establish a
direct contact with the experimental observations, the next step of our
program will be to include effects of hydrodynamical expansion of the QGP as
well as an assessment of photon production from a hadronic gas with a finite
lifetime, from hadronization to freeze-out. Work in this direction is in
progress.

{\em Acknowledgements}. S.-Y.W.\ gratefully acknowledges the Andrew Mellon
Foundation for partial support. This work was supported by the NSF under
grants PHY-9605186 and PHY-9988720.

%%%%%%%%%%%%%%%% here goes the bibliography %%%%%%%%%%%
%

%
%%%%%%%%%%%%%%%% here ends the bibliography %%%%%%%%%%%

%%%%%%%%%%%%%%% figure 1 %%%%%%%%%%%%%%%%%%%%%%%%%%%%%%%%%%%%%%%%%%%%
%%%%%%%%%%%%%%%%%% end figure 1 %%%%%%%%%%%%%%%%%%%%%%%%%%%%%%%%%%%%%%

%%%%%%%%%%%%%%% figure 2 %%%%%%%%%%%%%%%%%%%%%%%%%%%%%%%%%%%%%%%%%%%%
%%%%%%%%%%%%%%%%%% end figure 2 %%%%%%%%%%%%%%%%%%%%%%%%%%%%%%%%%%%%%%

%%%%%%%%%%%%%%% figure 3 %%%%%%%%%%%%%%%%%%%%%%%%%%%%%%%%%%%%%%%%%%%%
%%%%%%%%%%%%%%%%%% end figure 3 %%%%%%%%%%%%%%%%%%%%%%%%%%%%%%%%%%%%%%
\end{multicols}
\end{document}